\documentstyle[epsfig,aps]{revtex}
\newcommand{\beq}{\begin{equation}}
\newcommand{\eeq}{\end{equation}}
\def\beqa{\begin{eqnarray}}
\def\eeqa{\end{eqnarray}}

\def\lap{\lower.5ex\hbox{$\; \buildrel < \over \sim \;$}}
\def\gap{\lower.5ex\hbox{$\; \buildrel > \over \sim \;$}}

\font\tenbb=msbm10
\font\sevenbb=msbm7
\font\fivebb=msbm5
\newfam\bbfam
\textfont\bbfam=\tenbb \scriptfont\bbfam=\sevenbb
\scriptscriptfont\bbfam=\fivebb

\begin{document}
\draft

\twocolumn[\hsize\textwidth\columnwidth\hsize\csname
@twocolumnfalse\endcsname
\preprint{IHES/P/00/??\\  December 2000}

\title{E$_{10}$, BE$_{10}$ and Arithmetical Chaos in
Superstring Cosmology}

\author{Thibault Damour$^{1)}$ and Marc Henneaux$^{2)}$}

\address{$^{1)}$Institut des Hautes Etudes Scientifiques,
35, route de Chartres, F-91440 Bures-sur-Yvette, France \\
$^{2)}$Physique Th\'eorique et Math\'ematique, Universit\'e
Libre de Bruxelles, C.P.231, B-1050, Bruxelles, Belgium}

\maketitle
\begin{abstract}
It is shown that the never ending oscillatory behaviour of the generic solution, near a
cosmological singularity, of the massless bosonic sector of superstring theory can
be described as a billiard motion within a simplex in 9-dimensional hyperbolic space. The Coxeter
group of reflections of this billiard is discrete and is the Weyl group of the hyperbolic
Kac-Moody algebra E$_{10}$ (for type II) or BE$_{10}$ (for type I or heterotic),
 which are both arithmetic. These
results lead to a proof of the chaotic (``Anosov'') nature of the classical cosmological
oscillations, and suggest a ``chaotic quantum billiard'' scenario of vacuum selection in string
theory.
\end{abstract}
 \vskip1pc]
 One crucial problem in
string theory is the problem of vacuum selection. It is reasonable to believe that this problem
can be solved only in the context of cosmology, by studying the time evolution of {\it generic},
inhomogeneous (non-SUSY) string vacua. 
In this vein, it has been recently found \cite{DH1,DH2} that the general
solution near a spacelike singularity $(t \rightarrow 0)$ of the massless bosonic sector of all
superstring models ($D = 10$ IIA, IIB, I, HE, HO), as well as that of $M$-theory ($D = 11$ SUGRA),
exhibits a never ending oscillatory behaviour of the Belinskii-Khalatnikov-Lifshitz (BKL) type
\cite{BKL}. In this letter, we analyze in more detail the asymptotic dynamics (as $t \rightarrow
0$) of this oscillatory behaviour.  We find that the evolution of the scale factors and the dilaton
at each
spatial point can be be viewed as a billiard motion in some simplices in hyperbolic space $H^9$, which
have remarkable connections with hyperbolic Kac-Moody algebras of rank $10$. 

The central idea of the BKL approach is that the various points in space approximately decouple as
one approaches a spacelike singularity $(t \rightarrow 0)$. More precisely, the partial
differential equations that control the time evolution of the fields 
can be replaced by ordinary differential equations with
respect to time, with coefficients that are (relatively) slowly varying in space and time.
The details of how this is done are explained in \cite{BKL} for pure gravity, and in \cite{DH1,DH2}
for the graviton-dilaton-$p$-form systems relevant to superstring/$M$-theory.

We shall focus in this letter on the dynamical behaviour of the metric and the string dilaton and
recall first the relevant equations from \cite{DH1,DH2}.  To leading order, the metric (in
either the Einstein frame or the string frame) reads $g_{\mu \nu} \, dx^{\mu} \, dx^{\nu} = -N^2
(dx^0)^2 + \sum_{i=1}^{d} \ a_i^2 \, (\omega^i)^2$, where $d \equiv D-1$ denotes the spatial
dimension, and where $\omega^i (x) = e_j^i (x) \, dx^j$ is a $d$-bein whose time-dependence is
neglected compared to that of the local scale factors $a_i$.  It is convenient to work with the
10 field variables $\beta^{\mu}, \;  \mu = 1,\ldots, 10$, with, in the superstring 
(Einstein-frame) case,
$\beta^i \equiv -\ln \, a_i$ ($i = 1, \ldots, 9$), and $\beta^{10} \equiv -\varphi$ where $\varphi$ is 
the Einstein-frame dilaton. 
[In $M$-theory there is no dilaton but $\mu \equiv i =  1, \ldots, 10$. In the string
frame, we define $\beta_S^0 \equiv -\ln ( \sqrt {g_S} e^{-2 \Phi})$
and label $\mu = 0, \ldots, 9$.] 

We consider the evolution near a past (big-bang) or future (big-crunch) 
spacelike singularity located at $t=0$, where $t$ is the
proper time from the singularity. In the gauge $N = -\sqrt g$ (where $g$ is the
determinant of the Einstein-frame spatial metric), 
i.e. in terms of the new time variable $ d\tau = - dt / \sqrt
g$, the action (per unit comoving volume) describing the asymptotic dynamics of $\beta^{\mu}$ as
$t \rightarrow 0^+$ or $\tau \rightarrow + \infty$ has the form
\begin{equation}
S = \int d \tau \left[ G_{\mu \nu} \, \frac{d \beta^{\mu}}{d\tau} \ \frac{%
d\beta^{\nu}}{d\tau} - V (\beta^{\mu}) \right] \, ,  \label{eq1}
\end{equation}
\begin{equation}
V (\beta) \simeq \sum_A \, C_A \, e^{ - 2 w_A (\beta)} \, .  \label{eq2}
\end{equation}
In addition, the time reparametrization invariance (i.e. the equation of motion of $N$ in a
general gauge) imposes the usual ``zero-energy" constraint
$E = G_{\mu \nu} (d\beta^{\mu}/{d\tau}) (d\beta^{\nu}/{d\tau}) +
V (\beta^{\mu}) = 0$.
The metric $G_{\mu \nu}$ in field-space is a 10-dimensional metric of Lorentzian signature $-++
\cdots +$. Its explicit expression depends on the model and the choice of variables. In
$M$-theory, $ G_{\mu \nu}^M \, d\beta_M^{\mu} \, d\beta_M^{\nu} = \sum_{\mu=1}^{10} \,
(d\beta_M^{\mu})^2 - \left( \sum_{\mu=1}^{10} \, d\beta_M^{\mu} \right)^2$,
while in the string models, 
 $ G_{\mu \nu}^S \, d\beta_S^{\mu} \, d\beta_S^{\nu} = \sum_{i=1}^{9} \, (d\beta_S^i)^2
- (d\beta_S^0)^2$ in the string frame.
 Each exponential term, labelled by $A$, in the potential
$V (\beta^{\mu}),$ Eq.~(\ref {eq2}),
represents the effect, on the evolution of $(g_{\mu \nu} , \varphi)$, of either (i) the spatial
curvature of $g_{ij}$ (``gravitational walls''), (ii) the energy density of some electric-type
components of some $p$-form $A_{\mu_1 \ldots \mu_p}$ (``electric $p$-form
wall''), or (iii) the energy density of some magnetic-type components of $%
A_{\mu_1 \ldots \mu_p}$ (``magnetic $p$-form wall''). The coefficients $C_A$ are all found to be
positive, so that all the exponential walls in Eq.~(\ref {eq2}) are repulsive. The $C_A$'s vary in
space and time, but we neglect their variation compared to the asymptotic effect of $w_A (\beta)$
discussed below. Each exponent $ - 2 \, w_A (\beta)$ appearing in Eq.~(\ref{eq2}) is a {\it linear
form} in the $\beta^{\mu} : w_A (\beta) = w_{A\mu} \, \beta^{\mu} $.  The complete list of ``wall
forms'' $w_A (\beta)$, was given in \cite{DH1} for each string model.  The number of walls is
enormous, typically of the order of $700$.

At this stage, one sees that the $\tau$-time dynamics of the variables $ \beta^{\mu}$ is described
by a Toda-like system in a Lorentzian space, with a zero-energy constraint. But it seems
daunting to have to deal with $\sim 700$ exponential walls! However, the problem can be greatly
simplified because many of the walls turn out to be asymptotically irrelevant. To see this, it
is useful to project the motion of the variables $\beta^{\mu}$ onto
the 9-dimensional hyperbolic space $H^9$ (with curvature $-1$). This can be done because the
motion of $\beta^{\mu}$ is always time-like, so that, starting (in our units) from the origin, it
will remain within the 10-dimensional Lorentzian light cone of $G_{\mu \nu}$. This follows from
the energy constraint and the positivity of $V$ .  With our definitions,
 the evolution occurs in the {\it future} light-cone. The
projection to $H^9$ is performed by decomposing the motion of $\beta^{\mu}$ into its radial and
angular parts (see \cite{Chitre,Misner} and the generalization \cite{Melnikov}; see also
\cite{foot} for recent comments on the covariance of the chaos 
obtained in the billiard approximation). One writes
 $\beta^{\mu} =  + \rho \, \gamma^{\mu}$ with $\rho^2 \equiv - G_{\mu \nu} \,
\beta^{\mu} \, \beta^{\nu}$, $\rho >0$ and $G_{\mu \nu} \, \gamma^{\mu} \, \gamma^{\nu} = -1$ (so
that $\gamma^{\mu}$ runs over $H^9$, realized as the {\it future}, unit hyperboloid) and one introduces
a new evolution parameter: $dT =  k \, d\tau / \rho^2$. The action (\ref{eq1}) becomes
\begin{equation}
S = k \int dT \left[ - \left( \frac{d \, \ln \rho}{dT} \right)^2 + \left(
\frac{d \mbox{\boldmath$\gamma$}}{dT} \right)^2 - V_T (\rho , %
\mbox{\boldmath$\gamma$}) \right]  \label{eq6}
\end{equation}
where $d \mbox{\boldmath$\gamma$}^2 = G_{\mu \nu} \, d \gamma^{\mu} \, d \gamma^{\nu}$ is the
metric on $H^9$, and where $V_T = k^{-2} \, \rho^2 \, V = \sum_A \ k^{-2} \, C_A
\, \rho^2 \, \exp (-2 \, \rho \, w_A (\mbox{\boldmath$\gamma$}))$. When $t \rightarrow
0^+$, i.e. $\rho \rightarrow +\infty$, the transformed potential $V_T (\rho ,
\mbox{\boldmath$\gamma$})$ becomes sharper and sharper and reduces in the limit to a set of
$\rho$-independent impenetrable walls located at $w_A(\mbox{\boldmath$\gamma$}) = 0$
 on the unit hyperboloid (i.e. $V_T = 0$ when $w_A(\mbox{\boldmath$\gamma$}) > 0$,
and $V_T = +\infty$ when $w_A(\mbox{\boldmath$\gamma$}) < 0$). 
In this limit, $d \, \ln \rho / dT$ becomes constant, and one can
choose the constant $k$ so that $d \, \ln \rho / dT = 1 $.  The (approximately) linear motion 
of $\beta^{\mu}(\tau)$ between two ``collisions'' with the original multi-exponential potential
 $V (\beta^{\mu})$ is thereby mapped onto a geodesic motion of $\mbox{\boldmath$\gamma$} (T)$
  on $H^9$, interrupted by specular collisions on sharp hyperplanar walls.
   This motion has unit velocity $(d \mbox{\boldmath$\gamma$}
/ dT)^2 = 1$ because of the energy constraint.  In
terms of the original variables $\beta^{\mu}$, the motion is confined to the convex domain 
(a cone in
a 10-dimensional Minkowski space) defined by the intersection of the {\it positive} sides of all the
wall hyperplanes $w_A (\beta) = 0$ and of the interior of the
 future light-cone $G_{\mu \nu} \, \beta^{\mu} \,\beta^{\nu} = 0$.

A further, useful simplification is obtained by quotienting the dynamics of $\beta^{\mu}$ by the
natural permutation symmetries inherent in the problem, which correspond to ``large
diffeomorphisms" exchanging the various proper directions of expansion and the corresponding
scale factors. The natural configuration
space is therefore ${\fam\bbfam R}^d / {\rm S}_d$, which can be parametrized by the {\it ordered
multiplets} $ \beta^1 \leq \beta^2 \leq \cdots \leq \beta^d$. 
This quotienting is standard in most investigations of the BKL oscillations \cite{BKL}
and can be implemented in ${\fam\bbfam R}^d$ by introducing further sharp walls located at
$\beta^i = \beta^{i+1}$.  Note that the natural permutation symmetry group is different in
$M$-theory (where it is S$_{10}$), and in the $D = 10$ string models (S$_9$), and would be still
smaller in the successive dimensional reductions of these theories. However, there is a natural
consistency in quotienting each model by its natural permutation symmetry. Indeed, one finds that,
upon dimensional reduction, there arise new (exponential) walls, which replace the missing
permutation symmetries in lower dimensions \cite{DH3}. Finally the
dynamics of the models is equivalent, at each spatial point, to a hyperbolic billiard problem. The
specific shape of this model-dependent billiard is determined by the original walls and 
the permutation walls. Only the ``innermost'' walls (those which are not ``hidden'' behind others)
are relevant.

We have determined the set of innermost walls for all string models. The analysis is 
straightforward \cite{DH3} and we
report here only the final results, which are remarkably simple. Instead of the ${\cal O} (700)$
original walls we find, in all cases, that there are only 10 relevant walls. In fact, the seven
string theories M, IIA, IIB, I, HO, HE and the closed bosonic string in $D=10$ \cite{boso}, split
into {\it three} separate blocks of theories, corresponding to three distinct billiards.
 The first block (with 2 SUSY's in $D = 10$) is ${\cal B}_2 = \{$M, IIA, IIB$\}$ and
its ten walls are (in the natural variables of $M$-theory $\beta^{\mu} = \beta_M^{\mu}$),
\begin{eqnarray}
{\cal B}_2: w_i^{[2]} (\beta) &=& - \beta^i + \beta^{i+1} ( i = 1, \ldots , 9), \nonumber \\
w_{10}^{[2]} (\beta) &=& \beta^1 + \beta^2 + \beta^3.
\end{eqnarray}
The second block is ${\cal B}_1 = \{$I, HO, HE$\}$ and its ten walls read (when written in terms
of the string-frame variables of the heterotic theory $\alpha^i = \beta_S^i$, $\alpha^0 =
\beta_S^0$)
\begin{eqnarray}
{\cal B}_1: w_1^{[1]} (\alpha) &=& \alpha^1, \; w_i^{[1]} (\alpha) = - \alpha^{i-1} + \alpha^i (i
= 2
, \ldots , 9), \nonumber \\
w_{10}^{[1]} (\alpha) &=& \alpha^0 - \alpha^7 - \alpha^8 - \alpha^9.
\end{eqnarray}
The third block is simply ${\cal B}_0 = \{ D = 10$ closed bosonic$\}$ and its ten walls read (in string
variables) 
\begin{eqnarray}
{\cal B}_0: w_1^{[0]} (\alpha) &=& \alpha^1 + \alpha^2, \;  w_i^{[0]} (\alpha) = -
\alpha^{i-1} + \alpha^i (i = 2, \ldots , 9), \nonumber \\
  w_{10}^{[0]} (\alpha) &=& \alpha^0 - \alpha^7 -\alpha^8 - \alpha^9.
\end{eqnarray}
 In all cases, these walls define a simplex of $H^9$ which is non-compact but
of finite volume, and which has remarkable symmetry properties. 

\begin{figure}
\epsfig{file = 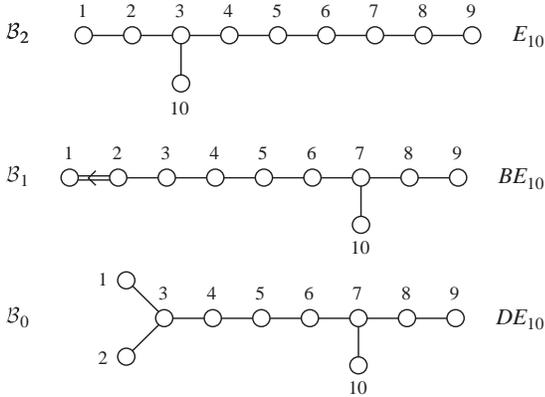,width = 0.4\textwidth, height =
0.3\textwidth,
angle = 0}
\caption{Dynkin diagrams defined (for each $n = 2,1,0$) by the ten wall forms
 $w_i^{[n]}(\beta^{\mu}), i = 1,\ldots,10$ that determine the billiard dynamics, near a
cosmological singularity, of the
three blocks of theories ${\cal B}_2 = \{$M, IIA, IIB$\}$, ${\cal B}_1 = \{$I, HO, HE$\}$ and
${\cal B}_0 = \{D = 10$ closed bosonic$\}$. The node labels $ 1,\ldots,10$ correspond to
the form label $i$ used in the text.
} \label{fig1}
\end{figure}

The most economical way to describe the geometry of the simplices is through their Coxeter
diagrams. This diagram encodes the angles between the faces and is obtained by computing the Gram
matrix of the scalar products between the unit normals to the faces, say $\Gamma_{ij}^{[n]} \equiv
\widehat{w}_i^{[n]} \cdot \widehat{w}_j^{[n]}$ where $\widehat{w}_i \equiv w_i / \sqrt{w_i \cdot
w_i}$, $i = 1 , \ldots , 10 $ labels the forms defining the (hyperplanar) faces of a simplex, and
the dot denotes the scalar product (between co-vectors) induced by the metric $ G_{\mu \nu} : w_i
\cdot w_j \equiv G^{\mu \nu} \, w_{i\mu} \, w_{j\nu}$ for $ w_i (\beta) = w_{i\mu} \,
\beta^{\mu}$. This Gram matrix does not depend on the normalization of the forms $w_i$ but
actually, all the wall forms $w_i$ listed above are normalized in
a natural way, i.e. have a natural length.
This is clear for the forms which are directly associated with dynamical walls in $D = 10$ or 11,
but this can also be extended to all the permutation-symmetry walls because
 they appear as dynamical walls after dimensional reduction \cite{DH3}. When the wall
forms are normalized accordingly (i.e. such that $V_i^{{\rm dynamical}} \propto \exp (-2 \, w_i
(\beta)$), they all have a squared length $w_i^{[n]} \cdot w_i^{[n]} = 2$, {\it except} for $
w_1^{[1]} \cdot w_1^{[1]} = 1$ in the ${\cal B}_1$ block. We can then compute
 the ``Cartan matrix'', $ a_{ij}^{[n]} \equiv 2 \, w_i^{[n]} \cdot
w_j^{[n]} / w_i^{[n]} \cdot w_i^{[n]}$, and the corresponding Dynkin diagram. 
One finds the diagrams given in Fig.~\ref{fig1}.

  The corresponding Coxeter diagrams are obtained
from the Dynkin diagrams by forgetting about the norms of the wall forms, i.e., by deleting the
arrow in $BE_{10}$. As can be seen from the figure, the Dynkin diagrams associated with the
billiards turn out to be the Dynkin diagrams of the following rank-$10$ hyperbolic Kac-Moody
algebras (see \cite {Kac}): E$_{10}$, BE$_{10}$ and DE$_{10}$ (for ${\cal B}_2$, ${\cal B}_1$ and
${\cal B}_0$, respectively). It is remarkable that the three billiards exhaust the only three
possible simplex Coxeter diagrams on $H^9$ with {\it discrete} associated Coxeter group (and this
is the highest dimension where such simplices exist) \cite {Vinberg}.
The analysis suggests to identify the 10 wall forms $w_i^{[n]} (\beta)$, $i = 1, \ldots , 10$ of
the billiards ${\cal B}_2$, ${\cal B}_1$ and ${\cal B}_0$ with a basis of simple roots of the
hyperbolic Kac-Moody algebras E$_{10}$, BE$_{10}$ and DE$_{10}$, while the 10 dynamical variables
$ \beta^{\mu}$, $\mu = 1 , \ldots , 10$, can be considered as parametrizing a generic vector in the
Cartan subalgebra of these algebras. It was conjectured some time ago \cite{Julia} that E$_{10}$
should be, in some sense, the symmetry group of SUGRA$_{11}$ reduced to one dimension 
(and that DE$_{10}$ be that of type I SUGRA$_{10}$, which has the same bosonic spectrum as
the bosonic string). Our
results, which indeed concern  the one-dimensional reduction, \`a la BKL, of $M$/string theories
 exhibit a clear sense in which E$_{10}$ lies behind the one-dimensional
evolution of the block ${\cal B}_2$ of theories: their asymptotic cosmological evolution as $ t
\rightarrow 0$ is a billiard motion, and the group of reflections in the walls of this billiard is
nothing else than the {\it Weyl group} of E$_{10}$ (i.e. the group of reflections in the
hyperplanes corresponding to the roots of E$_{10}$, which can be generated by the 10 simple roots
of its Dynkin diagram). It is intriguing -- and, to our knowledge, unanticipated 
(see, however, \cite{CJLP})-- that the
cosmological evolution of the second block of theories, ${\cal B}_1 = \{$I, HO, HE$\}$, be
described by {\it another} remarkable billiard, whose group of reflections is the Weyl group of
BE$_{10}$. The root lattices of E$_{10}$ and BE$_{10}$ exhaust the only two possible
unimodular even and odd Lorentzian 10-dimensional lattices \cite {Kac}.


A first consequence of the exceptional properties of the billiards concerns the nature of the
cosmological oscillatory behaviour. They lead to a direct technical proof that these oscillations,
for all three blocks, are chaotic in a mathematically well-defined sense. This is done
by reformulating, in a standard manner, the billiard dynamics as an equivalent collision-free
geodesic motion on a hyperbolic, finite-volume {\it manifold} (without boundary)
 ${\cal M}$ obtained by quotienting $H^9$
by an appropriate torsion-free discrete group. 
These geodesic motions define the ``most chaotic'' type of dynamical systems.
They are Anosov flows \cite{Anosov}, which imply, in particular, that they are ``mixing''. In
principle, one could (at least numerically) compute their largest, positive Lyapunov exponent, say
$\lambda^{[n]}$, and their (positive) Kolmogorov-Sinai entropy, say $h^{[n]}$. As we work on a
manifold with curvature normalized to $-1$, and walls given in terms of equations containing only
numbers of order unity, these quantities will also be of order unity. Furthermore, the two Coxeter
groups of E$_{10}$ and BE$_{10}$ are the only reflective {\it arithmetic} groups in 
$H^9$ \cite{Vinberg} so that the chaotic
motion in the fundamental simplices of E$_{10}$ and BE$_{10}$ will be of the exceptional
``arithmetical'' type \cite{BGGS}. We therefore expect that
 the quantum motion on these two billiards, and in particular the spectrum of
the Laplacian operator, exhibits exceptional features (Poisson statistics of
level-spacing,$\ldots$), linked to the existence of a Hecke
algebra of mutually commuting, conserved operators. Another (related) remarkable feature of the
billiard motions for all these blocks is their link, pointed out above,
 with Toda systems. This fact is probably quite significant, both classically and 
 quantum mechanically, because
Toda systems whose walls are given in terms of the simple roots of a Lie algebra enjoy 
remarkable properties. We leave to future work a study of our Toda systems which involve
infinite-dimensional hyperbolic Lie algebras.

The present investigation a priori concerned only the ``low-energy'' $(E \ll
(\alpha^{\prime})^{-1/2})$, classical cosmological behaviour of string
theories. In fact, if (when going toward the singularity) one starts at some
``initial'' time $t_0 \sim (d\beta / dt)_0^{-1}$ and
 insists on limiting the application of our results
to time scales $\vert t \vert \lower.5ex\hbox{$\; \buildrel > \over \sim \;$}
(\alpha^{\prime})^{1/2} \equiv t_s$, the total number of ``oscillations'',
i.e. the number of collisions on the walls of our billiard will be finite,
and will not be very large. The results above show that the number of
collisions between $t_0$ and $t \rightarrow 0$ is of order $N_{{\rm %
coll}} \sim \ln \tau \sim \ln (\ln (t_0/t))$. This is only $N_{{\rm coll}}
\sim 5$ if $t_0$ corresponds to the present Hubble scale and $t$ to the
string scale $t_s$. However, the
strongly mixing properties of geodesic motion on hyperbolic spaces make it
large enough for churning up the fabric of spacetime and transforming any,
non particularly homogeneous at time $t_0$, patch of space into a turbulent
foam at $t = t_{s}$. Indeed, the mere fact that the walls
associated with the spatial curvature and the form fields repeatedly rise up
(during the collisions) to the same level as the ``time'' curvature terms $%
\sim t^{-2}$, means that the spatial inhomogeneities at $t \sim t_s$ will also be of order
$t_s^{-2}$, corresponding to a string scale foam.

Our results on the ${\cal B}_2$ theories probably involve a deep (and not a priori evident)
connection with those of Ref.~\cite{BFM} on the structure of the moduli space of $M$-theory
compactified on the ten torus $T^{10}$, with vanishing 3-form potential. In both cases the Weyl
group of E$_{10}$ appears. In our case it is (partly) {\it dynamically} realized as reflections in
the walls of a billiard, while in Ref.~\cite{BFM} it is {\it kinematically} realized as a symmetry
group of the moduli space of compactifications preserving the maximal number of supersymmetries.
In particular, the crucial E-type node of the Dynkin diagram of E$_{10}$ (Fig.~\ref{fig1}) comes,
in our study and in the case of $M$-theory, from the wall form $w_{10}^{[2]} (\beta) = \beta_M^1 +
\beta_M^2 + \beta_M^3$ associated with the electric energy of the 3-form. By contrast, in
\cite{BFM} the 3-form is set to zero, and the reflection in $w_{10}^{[2]}$ comes from the 2/5
duality transformation (which is a double $T$ duality in type II theories), which exchanges (in
$M$-theory) the 2-brane and the 5-brane. As we emphasized above, dimensional reduction transforms
kinematical (permutation) walls into dynamical ones. This suggests that there is no difference of
nature between our walls, and that, viewed from a higher standpoint (12-dimension ?), they would
all look kinematical, as they are in \cite{BFM}. By analogy, our findings for the ${\cal B}_1$
theories suggest that the Weyl group of BE$_{10}$ is a symmetry group of the moduli space of
 $T^{9}$ compactifications of $\{$I, HO, HE$\}$.

Perhaps the most interesting aspect of the above analysis is to provide hints for a scenario of
vacuum selection in string cosmology. If we heuristically extend our (classical, low-energy,
tree-level) results to the quantum, stringy $(t \sim t_s)$ and/or strongly coupled $(g_s \sim 1)$
regime, we are led to conjecture that the initial state of the universe is equivalent to the
quantum motion in a certain {\it finite volume chaotic} billiard. This billiard is (as in a hall
of mirrors game) the fundamental polytope of a discrete symmetry group which contains, as
subgroups, the Weyl groups of both E$_{10}$ and BE$_{10}$ \cite{footnote}. We are here assuming
that there is (for finite spatial volume universes) a non-zero transition amplitude between the
moduli spaces of the two blocks of superstring ``theories'' (viewed as ``states'' of an underlying
theory). If we had a description of the resulting combined moduli space (orbifolded by its
discrete symmetry group) we might even consider as most probable initial state of the
universe the fundamental mode of the combined billiard, though this does not seem crucial
for vacuum
selection purposes. This picture is a generalization of the picture of Ref.~%
\cite{HM} and, like the latter, might solve the problem of cosmological
vacuum selection in allowing the initial state to have a finite probability
of exploring the subregions of moduli space which have a chance of inflating
and evolving into our present universe.

We thank Fran\c cois Bonahon, Costas Bachas, Michael Douglas,
 Fran\c cois Labourie, Nikita Nekrasov, Hermann Nicolai and,
especially, Victor Kac, for informative discussions.


\begin{references}
\bibitem{boso} We include for the sake of comparison the bosonic string model in $D=10$.
Though $D = 10$ is not the critical dimension of the quantum bosonic string, it is
the critical dimension above which the never ending oscillations in its cosmological evolution
disappear \cite{DH2}.
\end{references}
\end{document}